\documentstyle[seceq,epsf,twoside]{ptptex}
\notypesetlogo  %comment in if to eliminate PTPTeX logo
\title{The $\sigma$-Meson Production in 
        Excited $\Upsilon$ Decay Processes}
\author{%
Toshihiko {\sc Komada},
Muneyuki {\sc Ishida}$^*$ and Shin {\sc Ishida}$^{**}$  }
\inst{%
Department of Engineering Science, Junior College Funabashi Campus,
           Nihon University, Funabashi 274-8501, Japan\\
$^*$Department of Physics, Tokyo Institute of Technology\\
Tokyo 152-8551, Japan\\
$^{**}$Atomic Energy Research Institute, 
College of Science and Technology\\
Nihon University, Tokyo 101-0062, Japan
}
\recdate{%
\today
}
\abst{%
We analyze 
the $\pi\pi$ production amplitudes in the excited $\Upsilon$ decay 
processes, 
$\Upsilon (2S)\to\Upsilon (1S)\pi^+\pi^-$,  
$\Upsilon (3S)\to\Upsilon (1S)\pi^+\pi^-$ and
$\Upsilon (3S)\to\Upsilon (2S)\pi^+\pi^-$,
and the $\pi\pi$ and $K\bar K$ production amplitudes in the 
charmonium decay processes,
$\psi (2S)\to J/\psi \pi^+\pi^-$ and 
$J/\psi\to \phi \pi^+\pi^- ,\ \phi K^+K^-$,
including the possible effect of light $\sigma$ production.
The amplitudes are parametrized by the sum of Breit-Wigner amplitudes
for the $\sigma$ and the other relevant particles 
and of the direct $2\pi$-production amplitude, 
following the VMW method.  
All the $\pi\pi$ (and $K\bar K$) mass spectra are reproduced 
well with the obtained values of $\sigma$-parameters,
$m_\sigma =526^{+48}_{-37} $MeV and $\Gamma_\sigma =301^{+145}_{-100} $MeV,
which is almost consistent with the values in 
our previous phase shift analyses.
}
\begin{document}
\maketitle

\setcounter{tocdepth}{4}

\section{Introduction}
Whether the light $\sigma$-meson really
exists or does not is an important problem in hadron physics.
For many years, its existence 
had been neglected phenomenologically 
mainly due to the negative result of the
analyses of $I=0$ $S$-wave $\pi\pi$ scattering phase shift. 

In many $\pi\pi$-production experiments, a  
large event concentration or a bump structure 
in the spectra of $\pi\pi$ invariant mass 
$m_{\pi\pi}$ around 500 MeV had been observed, 
however, conventionally it was not regarded as
$\sigma$-resonance, but as a mere $\pi\pi$-background, 
under influence of the
so callled ``universality argument."\cite{rf1} 
In this argument, it is stated that 
because of the unitarity of $S$-matrix
and of the analyticity of the amplitudes,
the $\pi\pi$ production amplitude ${\cal F}$ takes the form
${\cal F}=\alpha (s){\cal T}$ (${\cal T}$ being the $\pi\pi$ scattering
amplitude),
with a slowly varying real function $\alpha (s)$.
The pole position of $S$-matrix is determined solely
through the analysis of ${\cal T}$, which was believed to have 
no light $\sigma$-pole at that time.

Recently the data of $\pi\pi$-scattering phase shift 
had been reanalyzed by many groups\cite{rf2}
including ours
%\cite{rf3} 
and the existence of light $\sigma (450\sim 600)$
was strongly suggested.
The result of no $\sigma$-existence in the conventional analyses was 
pointed out\cite{rf4} to be due to the lack of consideration on the
cancellation mechanism guaranteed by chiral symmetry, 
and shown to be not correct. 
Furthermore, we have pointed out that 
the ``universality argument" should be revised, taking into account 
the quark physical picture of hadrons\cite{rf5}: 
The essential point is that the strong interaction is a residual interaction of QCD among
all color-neutral bound states of quarks, anti-quarks and gluons, and accordingly 
the requirement of unitarity and analyticity 
must be made on the $S$-matrix elements with the right 
``quark physical bases," including not only the stable $\pi$ meson 
but also the ``stable" $\sigma$ meson (which is also considered to be a $q\bar q$
bound state). 
The universality argument (and the conventional application of final
state interaction (FSI) theorem) 
with the $S$-matrix bases including only stable particles, is not correct:
Since this $\sigma$-meson has, in principle, an independent production 
coupling generally with a strong phase 
from the $2\pi$ system.

%%%%%%%
Thus, the production amplitude ${\cal F}$ has ``independent" properties
from the scattering amplitude ${\cal T}$, except that the pole position 
of resonant particles should be universal through both amplitudes.
Accordingly, the data of many $\pi\pi$-production experiments,
which had been analyzed following the universality argument, must be reanalyzed
independently from ${\cal T}$ 
by including the effect of $\sigma$-production.
We parametrize ${\cal F}$ as a sum of 
Breit-Wigner amplitudes for the relevant resonances 
(including $\sigma$-resonance) and of the direct $2\pi$ production 
amplitude, following the VMW method.
In the VMW method the above mentioned universal pole singularity is 
explicitly taken into account, and this method is consistent with the 
unitarity of the total $S$-matrix (and also with the FSI theorem, 
applied correctly) with the above metioned right bases.
The VMW method had already been applied
%\cite{rf6} to the $pp$-central collision $pp\to \pi\pi$, 
to the processes of $J/\psi\to\omega\pi\pi$ decay, 
$p\bar p\to 3\pi^0$ annihilation by us\cite{rfppbar}, and
essentially similar method to the process
$D^-\rightarrow \pi^-\pi^+\pi^+$ by E791\cite{gobel}.
The large event concentration in low $m_{\pi^+\pi^-}$ region 
in $J/\psi\to\omega\pi\pi$ and $D^-\rightarrow \pi^-\pi^+\pi^+$ was 
satisfactorily explained by $\sigma$-production,
while in $p\bar p\to 3\pi^0$ clear evidences for $\sigma$-meson 
prodution were shown. 
In this paper we apply this method to the analyses of 
the hadronic decays of excited $\Upsilon$,
$\Upsilon (2S,3S)$, and $\psi (1S,2S)$.\\ 
%%%%%%%%%%%%%%%%%%%%%%%%%%%%%%%%%%

\section{Method of the analyses}
We analyze\cite{rf0} the $m_{\pi\pi}$ spectra of the processes,
$\Upsilon (2S)\to\Upsilon (1S)\pi^+\pi^-$,
%\cite{rf7,rf8,rf9,rf10}
  $\Upsilon (3S)\to\Upsilon (1S)\pi^+\pi^-$,
%\cite{rf7} 
$\Upsilon (3S)\to\Upsilon (2S)\pi^+\pi^-$
%\cite{rf7} 
and 
$\psi (2S)\to J/\psi \pi^+\pi^-$
%\cite{rf11} 
following the VMW method in one-channel form.
The ${\cal F}$ is given
by a coherent sum 
of $\sigma$-Breit-Wigner amplitude and 
of direct $2\pi$-production amplitude as
\begin{eqnarray}
{\cal F } &=& \frac{e^{i\theta_\sigma}r_\sigma }{m_\sigma^2 -s - i\sqrt{s}\Gamma_\sigma (s)}
+r_{2\pi}e^{i\theta_{2\pi}};\nonumber \\
 & & \Gamma_\sigma (s) = \frac{g_\sigma^2 p_1(s)}{8\pi s}, \ \ 
p_1(s)=\sqrt{\frac{s}{4}-m_\pi^2} ,
\label{eq1}
\end{eqnarray}
where $r_\sigma (r_{2\pi })$ is the $\sigma (2\pi )$ production coupling and 
$\theta_\sigma (\theta_{2\pi})$ corresponds to the production phase.
These parameters are process-dependent. 
%%%%%%%%%%%%
The $J/\psi\to \phi \pi^+\pi^- ,\ \phi K^+K^-$ process
%\cite{rf12} 
is analyzed by VMW method in two-channel
form, where 
%the $\pi\pi$ and $KK$ production amplitudes, ${\cal F}_{\pi\pi}$ and ${\cal F}_{KK}$,
%respectively, are given by
%\begin{eqnarray}
%{\cal F}_{\pi\pi} &=& 
%\frac{e^{i\theta_\sigma}r_\sigma }{m_\sigma^2 -s - i\sqrt{s}\Gamma_\sigma (s)}+\sum_{f_0} 
%\frac{e^{i\theta_{f_0}}r_{f_0}g_{f_0\pi\pi} }{m_{f_0}^2 -s - i\sqrt{s}\Gamma_{f_0}^{\rm tot} (s)}
%+r_{2\pi}e^{i\theta_{2\pi}};  \nonumber\\
%{\cal F}_{KK} &=& \sum_{f_0} \frac{e^{i\theta_{f_0}}r_{f_0}g_{f_0KK} }{m_{f_0}^2 -s - i\sqrt{s}\Gamma_{f_0}^{\rm tot} (s)}
%+r_{2K}e^{i\theta_{2K}};\ \ \ 
%f_0=
$f_0(980),\ f_0(1370),\ f_0(1500)$,
%\nonumber\\
% & &  
%\Gamma_{f_0}^{\rm tot} = \Gamma_{f_0}^{\pi\pi}+\Gamma_{f_0}^{KK}
%         =\frac{p_1 g_{f_0\pi\pi}^2}{8\pi s} + \frac{p_2 g_{f_0KK}^2}{8\pi s}.
% \nonumber
%\end{eqnarray}
as well as $\sigma$, are included. 
%
%By using the above formulas of production amplitudes 
%the $m_{\pi\pi}$ or $m_{KK}$ spectra (that is, $\sqrt s$ spectra)
%are given by
%\begin{eqnarray} 
%\frac{d \sigma}{d\sqrt s} &=& \frac{p(M'^2;M,\sqrt s)p_1(s)}{32\pi^3 M'^2}|{\cal F}|^2 ,\ \ \ 
%p(M'^2;M,\sqrt s)=\frac{\sqrt{(M'^2-M^2-s)^2-4M^2 s}}{2\sqrt{s}}, \nonumber
%\end{eqnarray}
%where $M'(M)$ is the mass of the initial (final) quarkonium.
%
All the relevant spectra are fitted by using common values of the parameters, 
the mass of $m_\sigma$ and the $\sigma\pi\pi$ coupling $g_{\sigma\pi\pi}$.

\section{Results and Conclusion}

\begin{figure}[t]
  \epsfxsize=14 cm
  \epsfysize=16 cm
 \centerline{\epsffile{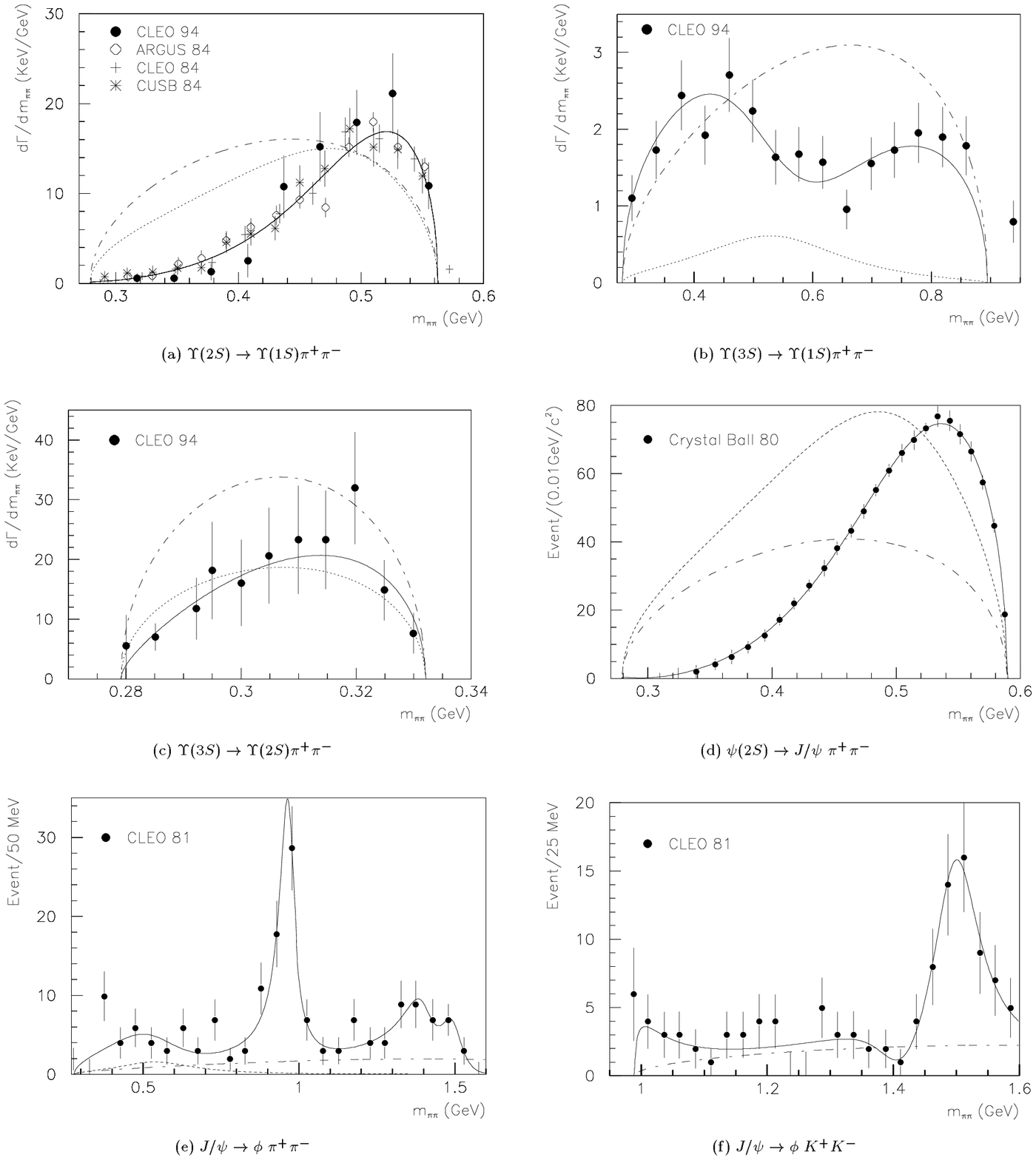}}
 \caption{The result of the fit to the $\pi\pi$ (or $KK$) mass spectra of
(a) $\Upsilon (2S)\to \Upsilon (1S) \pi\pi$, 
(b) $\Upsilon (3S)\to \Upsilon (1S) \pi\pi$, 
(c) $\Upsilon (3S)\to \Upsilon (2S) \pi\pi$, 
(d) $\psi (2S)\to J/\psi  \pi\pi$, 
(e) $J/\psi \to \phi \pi\pi$, and    
(f)  $J/\psi \to \phi KK$ .  
The bold line represents the fit, while the dotted (dot-dashed) does the contribution
of $\sigma$(direct $2\pi$)-production. 
By multiplying appropriate proportional factors,
the different data sets are adjusted to the one 
with the largest number of events.
%In (a) CLEO data\cite{rf7}, shown by black circle, with factor 1;
%ARGUS\cite{rf8} by open circle with factor 1.5; CLEO\cite{rf9} by cross with factor 2;
%CUSB\cite{rf10} by asterisk with factor 3.
%Data in (b), (c) and (d)  are respectively, 
%from CLEO\cite{rf7}, CLEO\cite{rf7} and Crystal Ball\cite{rf11}.
%Data in (e) and (f) are from MARK II\cite{rf12}.
For experimental references, see \cite{rf0}.
}
  \label{fig:1}
\end{figure}

The results of our analysis are shown in Fig.~1. 
The spectra of six different processes 
%((a) to (e) in Fig. 2) 
are reproduced well 
by the interference between the $\sigma$-Breit-Wigner
amplitude with 
%\begin{eqnarray}  
$m_\sigma = 526 ^{+48}_{-37}{\rm MeV}, \  
\Gamma_\sigma =301^{+145}_{-100}{\rm MeV}$,
%\label{eq3}
%\end{eqnarray}
and the direct $2\pi$ production amplitude.
The corresponding pole position is 
$\sqrt{s_{\rm pole}}=(535^{+48}_{-36})-i(155^{+76}_{-53})$MeV.
The total $\chi^2$ is $\chi^2/(N_{\rm data}-N_{\rm
param})=86.5/(150-37)=0.77$.
The contribution of $\sigma$ and of direct $2\pi$ production amplitudes to the spectra
are given, respectively, by dot and dot-dashed lines in this figure. 
It is notable that 
the destructive interference between $\sigma$ amplitude and $2\pi$ amplitudes 
explains the suppression of the spectra 
in the threshold region of $\Upsilon (2S\to 1S)$ and $\psi (2S\to 1S)$ decays,
while in $\Upsilon (3S\to 1S)$ decay these two amplitudes interfere constructively, 
and the steep increase from the threshold is reproduced.
These threshold behaviors of the production amplitudes are\cite{rf4} 
shown to be consistent with the restriction  
from chiral symmetry.
It is especially interesting that 
the double peak structure,
with the bottom around the $\sigma$-peak position,
of the spectra in $\Upsilon (3S\to 1S)$ decays
is also reproduced well 
by the above interference
between the direct $2\pi$ production amplitude with zero phase 
and the $\sigma$-production amplitude with a moving phase.
That is, we are observing the very phase motion of the $\sigma$-Breit-Wigner
formula through the variation of the spectra in this process.

The obtained values of masses and widths of 
$\sigma$-meson given above are almost 
consistent with the results obtained in our previous $\pi\pi$ 
phase shift analysis\cite{rf2},\\
$m_\sigma$=535$\sim$675MeV and
$\Gamma_\sigma$=385$\pm$70MeV. 
These results give strong evidence for existence of the light 
$\sigma$(450--600).

The consistency of our result of analyses with the general constraints 
from chiral symmetry is discussed in the separate talk\cite{PLB2}.\\

In this conference, there was raised a doubt\cite{ochs} 
on the $\sigma$-existence 
along the line of the universality argument;\\
i) on $J/\psi\rightarrow\omega\pi\pi$ process. 
The experimental cos$\theta$-distributions show the parabolic shape 
around cos$\theta\approx 0$ in the range of $m_{\pi\pi}$, 300 through 700 MeV,
coming from the interference between the $S$ and $D$ waves.
However, the reverse of the direction of the parabolic shape,
which is expected from the relative phase motion,
to the background $D$-wave phase, of the $\sigma$-Breit-Wigner amplitude,
was not observed. This criticism will become invalid, if we take into account the  
possible strong phase.  
For example, $\theta_\sigma$=$-90^\circ$.\\
ii) on $pp$-central collision, $pp\rightarrow pp\pi\pi$.
The result of the partial wave analysis by WA102 shows that 
the non-resonant $\pi\pi$-production from one pion exchange
is enough to reproduce the $S$- and $D$-wave components 
in low $m_{\pi\pi}$ region.
We consider that it is necessary, in addition to this,  
to consider the effect of 
the $\sigma$-Breit-Wigner amplitude,
since the relative phase between $S$- and $D$-wave amplitudes
shows a structure around $m_{\pi\pi}\sim 500$MeV region,
implying the interference between the non-resonant 
$r_{2\pi}e^{i\theta_{2\pi}}$ and $r_{\sigma}e^{i\theta_{\sigma}}$ 
in our formula Eq.~(\ref{eq1}).
By the way,
the experimental relative phase 
is completely different from that in the $\pi\pi$-scattering.  
This fact clearly shows the universality-argument 
(where $\alpha = {\cal F}/{\cal T}$ is supposed to be real) is not correct.\\
iii) on $\pi\pi$-scattering. No rapid phase variation due to $\sigma (600)$
has been observed.
We had mentioned in many occasions\cite{PLB2} the reason why 
the $\sigma$-Breit-Wigner phase motion is not directly observed in $\pi\pi$-scattering:  
Because of the constraints from chiral symmetry the $\sigma$-amplitude must be strongly
cancelled out by the non-resonant repulsive $\pi\pi$-amplitude in $\pi\pi$-scattering.

%\acknowledgements

%Thanks THANK. MANY THANKS!

\end{document}